# The cores of dwarf galaxy haloes

Julio F. Navarro[1,2], Vincent R. Eke[2], and Carlos S. Frenk[2]
[1] *Bart Bok Fellow. Steward Observatory, The University of Arizona, Tucson, AZ 85721, USA.*
[2] *Physics Department, University of Durham, Durham DH1 3LE, England.*

23 October 1996

**ABSTRACT**
We use N-body simulations to examine the effects of mass outflows on the density profiles of cold dark matter (CDM) haloes surrounding dwarf galaxies. In particular, we investigate the consequences of supernova-driven winds that expel a large fraction of the baryonic component from a dwarf galaxy disk after a vigorous episode of star formation. We show that this sudden loss of mass leads to the formation of a core in the dark matter density profile, although the original halo is modeled by a coreless (Hernquist) profile. The core radius thus created is a sensitive function of the mass and radius of the baryonic disk being blown up. The loss of a disk with mass and size consistent with primordial nucleosynthesis constraints and angular momentum considerations imprints a core radius which is only a small fraction of the original scale-length of the halo. These small perturbations are, however, enough to reconcile the rotation curves of dwarf irregulars with the density profiles of haloes formed in the standard CDM scenario.

## 1 INTRODUCTION

The luminous baryonic component (gas and stars) accounts for a very small fraction of the total virial mass measured within the luminous radius of dwarf irregular galaxies. As a result, measurements of the rotation curves of gas and stars in these systems yield valuable information about the inner structure of their surrounding dark haloes. A number of high-quality rotation curves for dwarf spirals and irregulars have become available in the past few years (see, for example, Carignan & Freeman 1988, Carignan & Beaulieu 1990, Broeils 1992a,b, Jobin & Carignan 1990, Lake, Schommer & van Gorkom 1990, Puche & Carignan 1991). Near the centre of these systems the circular velocity increases approximately linearly with radius, indicating that the surrounding dark haloes may have a constant density core of radius comparable to the luminous radius of the galaxy.

This finding is at odds with the results of N-body simulations of the formation of collisionless, non-baryonic dark haloes in hierarchical clustering theories. Haloes formed in scale-free universes or in a universe dominated by cold dark matter are found to have density profiles that appear to diverge near the centre as $r^{-1}$ and rule out constant density cores of the size required to match these observations (Dubinski & Carlberg 1991, Warren et al 1992, Crone, Evrard & Richstone 1994, Navarro, Frenk & White 1995, 1996). A fundamental reason underlies this result. A core with a well-defined central density implies a characteristic feature in the power spectrum of initial density fluctuations on scales where most favoured cosmological models would predict a nearly scale-free behaviour. (Hereafter we will reserve the term "core" to refer to a region of constant density near the centre.)

The apparent contradiction between the lack of cores in simulated CDM haloes and the rotation curves of dwarf galaxies has been highlighted in two recent papers (Flores & Primack 1994, Moore 1994). These authors argue that the *shape* of the density profiles of the simulated haloes is inconsistent with the rotation curves of DDO154, DDO168, DDO170 and NGC3109, four dwarf galaxies for which precision photometry as well as high signal-to-noise HI rotation curves are available. It is important to remember, however, that the centres of haloes are fragile structures that may very well be affected by the assembly of the luminous galaxy. Dwarf galaxies are extremely baryon-poor, a fact commonly attributed not to a baryon deficiency at the time of formation but rather to the effect of feedback from evolving stars and supernovae. The energy input from ageing stars is thought to drive large scale winds that expel a significant fraction of the original baryonic component out of these weakly bound systems, leaving behind the baryon-poor dwarfs that we observe today (Dekel & Silk 1986). It is unlikely that the central structure of the haloes could have survived this process unscathed. The sudden removal of the disk by winds can cause substantial damage to the central region of haloes, leading perhaps to the formation of a core. Since massive outflows are observed in a number of starburst galaxies (Heckman et al 1990), it is important to understand how this process affects the central properties of the surrounding halo.

In this paper, we use numerical simulations of a very simplified outflow model in order to assess whether sudden baryon losses can produce sizeable cores in initially coreless dark matter haloes. Our model uses N-body realizations of a Hernquist model (Hernquist 1990), a coreless dynamical model that fits rather well the structure of haloes formed in cosmological simulations. This model is perturbed with the potential of a disk that grows slowly in time in order to mimic the adiabatic changes in the halo structure caused by the assembly of a galactic disk. Once the halo has reached equilibrium we remove the disk potential and let the halo relax to a new equilibrium configuration. Although admittedly oversimplified, this model we believe captures the essential features of the collapse and outflow processes likely to occur during the formation of dwarf irregulars and can serve to assess their viability as a means to reconcile the results of cosmological N-body simulations with observations.

The plan of this paper is as follows. Section 2 describes our numerical experiments and presents the results of the simulations, while §3 compares our results with observations. Section 4 discusses the cosmological implications of our results. Our conclusions are summarized in §5.

## 2 NUMERICAL METHODS

### 2.1 The halo model

A convenient analytical potential-density pair that reproduces rather well the central structure of CDM haloes (Dubinski & Calberg 1991, Navarro et al 1995, 1996) has been proposed by Hernquist (1990). The density profile is given by

$$\rho_h(r) = \frac{M_h}{2\pi} \frac{a_h}{r(r+a_h)^3}, \quad (1)$$

where $M_h$ is the total mass of the system and $a_h$ is its scale length. The computational units of the model are fixed by assuming that $G = M_h = a_h = 1$. The circular velocity, $V_h(r) = (GM_h r)^{1/2}/(r+a_h)$, rises like $r^{1/2}$ from the centre, reaches a maximum $V_{max} = (GM_h/a_h)^{1/2}/2 = 0.5$ at $r = a_h = 1$, and decreases thereafter. The time to complete a circular orbit at $r = a_h$ is $4\pi a_h^{3/2}/(GM_h)^{1/2} = 4\pi$.

We construct a 10,000-particle realization of a spherically symmetric Hernquist model assuming isotropic orbits. The gravitational forces are Plummer-softened with $\epsilon = 0.03$. We test that the model is truly in equilibrium by letting it evolve in isolation for 200 time units. The evolution of the 10%, 25%, and 50% mass radial shells is shown in Figure 1 with dot-dashed lines. No significant evolution is seen in any of these shells, indicating that the model is initially in equilibrium.

### 2.2 The mass outflow model

The growth of a galactic disk inside a halo is modeled using an exponential disk external potential. This potential is fully specified by two parameters, the mass of the disk, $M_{disk}$, and its exponential scale-length, $r_{disk}$. In order to allow the halo to respond adiabatically to this potential we impose it over $\sim 20$ time units, letting the scale-length of the disk decrease linearly with time from 5 units to $r_{disk}$. We then let the disk+halo system evolve for $\sim 30$ time units, until the halo reaches equilibrium again. We checked explicitly that doubling these timescales has no appreciable effect on our results. Finally we model the sudden ejection of mass by removing the external disk potential altogether, and letting the halo relax to a new equilibrium configuration. We calculated models with three different values of the disk mass, $M_{disk} = 0.05, 0.1,$ and $0.2$, and three values of the disk scale-lengths, $r_{disk} = 0.01, 0.02,$ and $0.04$, for a total of nine runs. The perturbation to the halo is stronger for increasing values of the ratio $M_{disk}/r_{disk}$.

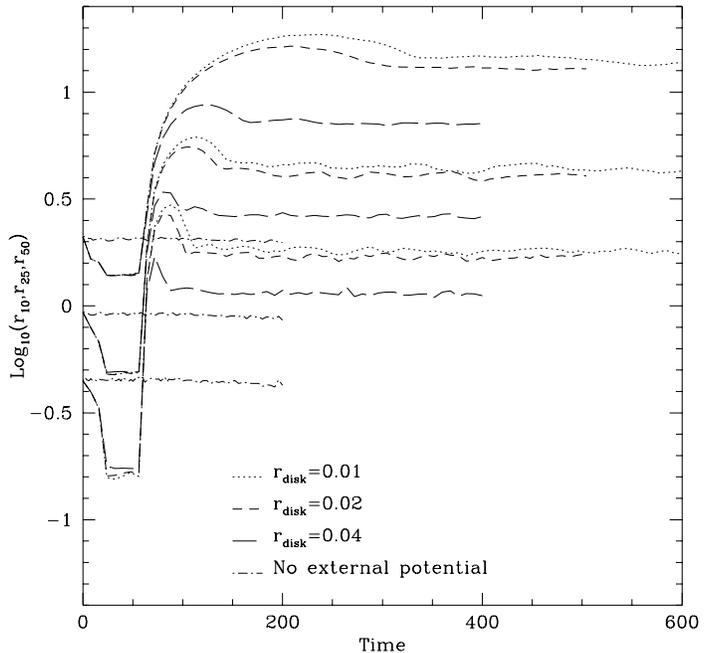

**Figure 1.** The evolution of radial mass shells in a Hernquist model run in isolation (dot-dashed lines), or subject to the collapse and removal of an exponential disk potential of mass $M_{disk} = 0.2$, and $r_{disk} = 0.01, 0.02,$ and $0.04$ (dotted, short-dashed and long-dashed, respectively).

### 2.3 Description of the runs.

Figure 1 shows the evolution of different halo mass shells in three runs with $M_{disk} = 0.2$ and $r_{disk} = 0.01, 0.02,$ and $0.04$, respectively. The halo is adiabatically compressed by the growing disk, settles into equilibrium once the growth of the disk is over, and expands suddenly after the disk potential is removed. Note that the central regions are the most affected by this process. Indeed, after the disk is removed many halo particles near the centre become unbound and leave the system. This is apparent in Figure 2, where we plot the phase-space (radius vs velocity) distribution of halo partices just before the disk potential is removed. Most particles near the centre have velocities larger than the escape velocity of the system computed neglecting the disk (solid line), and will therefore leave the system when the disk is removed.

Figures 3$a,b,c$ show the final density profile of the halo in our nine runs. The solid line shows the analytical Hernquist model (eq. 1), and the dot-dashed line is the Hernquist model evolved in isolation at $t = 200$. This figure shows clearly that swift mass outflows of the kind modeled here can lead to large modifications of the central structure of the haloes. The magnitude of the perturbation depends strongly on the mass and radius of the disk being removed. Figure 4 shows the core radii derived for each of the profiles in Figure 3 (computed by fitting a non-singular isothermal

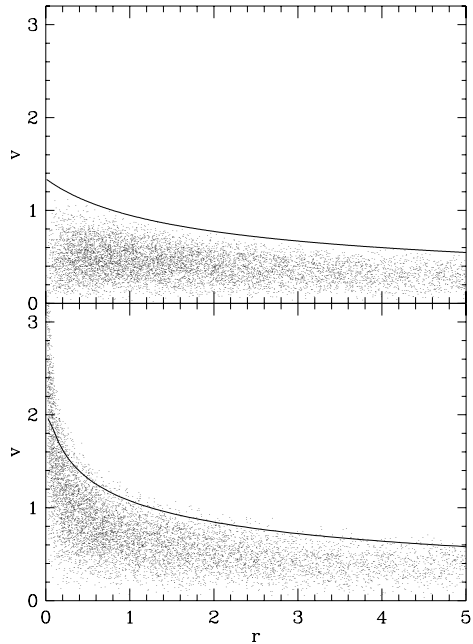

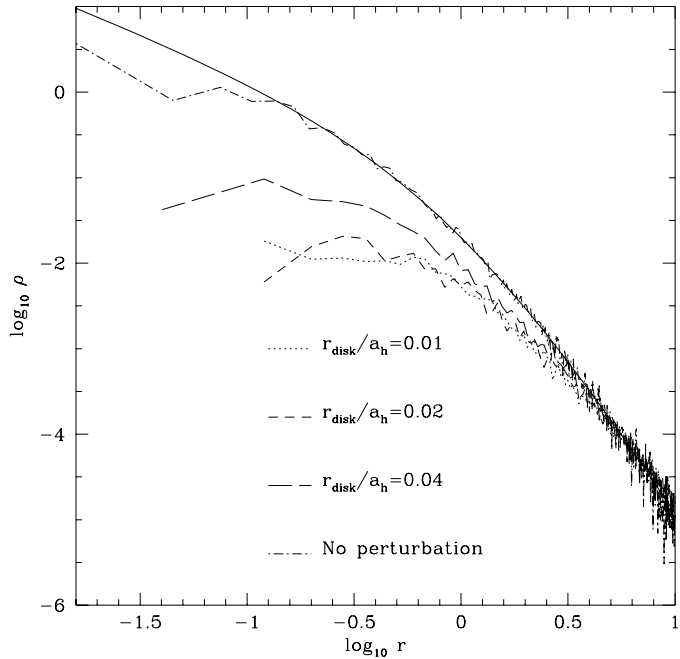

**Figure 2.** "Phase-space" distribution of halo particles before and after imposing the disk potential (upper and lower panel, repectively). The solid line in each panel shows the escape velocity corresponding to the halo before the addition of the disk. The solid line in the bottom panel shows the escape velocity of the halo after the perturbation, computed neglecting the contribution of the disk to the overall potential. Most particles above this line (preferentially those near the centre) will escape when the disk is removed.

**Figure 3.** (a) Equilibrium density profiles of haloes after removal of the disk ($M_{disk} = 0.2$). The solid line is the original Hernquist profile, common to all cases. The dot-dashed line is the equilibrium profile of the 10,000-particle realization of the Hernquist model run in isolation at $t = 200$.

profile to the inner 25% of the halo mass at the final time) as a function of the strength of the perturbation, measured by $M_{disk}/r_{disk}$. The data in this figure are well fitted by a simple relation, $r_{core} \approx 0.11(M_{disk}/r_{disk})^{1/2}$. The strongest perturbation ($M_{disk}/r_{disk} \sim 20$) produces a core radius as large as half the original scale-length of the disk. The weakest perturbation, ($M_{disk} = 0.05$ and $r_{disk} = 0.04$), results in a core radius which is about 15% of the original scale-length of the disk, only slightly larger than that imprinted on the halo by numerical inaccuracies in our modeling (see Figure 3c).

## 3 COMPARISON WITH OBSERVATIONS.

In this section we compare our results with the structure of dark matter haloes inferred from the rotation curves of four dwarf galaxies, DDO154 (Carignan & Beaulieu 1989), DDO168 (Broelis 1992), DDO170 (Lake *et al* 1990), and NGC3109 (Jobin & Carignan 1990). These galaxies are all dwarfs and baryon-poor, so the mechanism discussed in the previous section might have played an important role during their formation. The contribution of the dark halo to the circular velocity derived from these observations is shown in the panels of Figure 5 as solid lines. The two lines correspond to the "maximal disk" and "minimal disk" fits to the actual rotation curve quoted by the authors, and therefore bracket

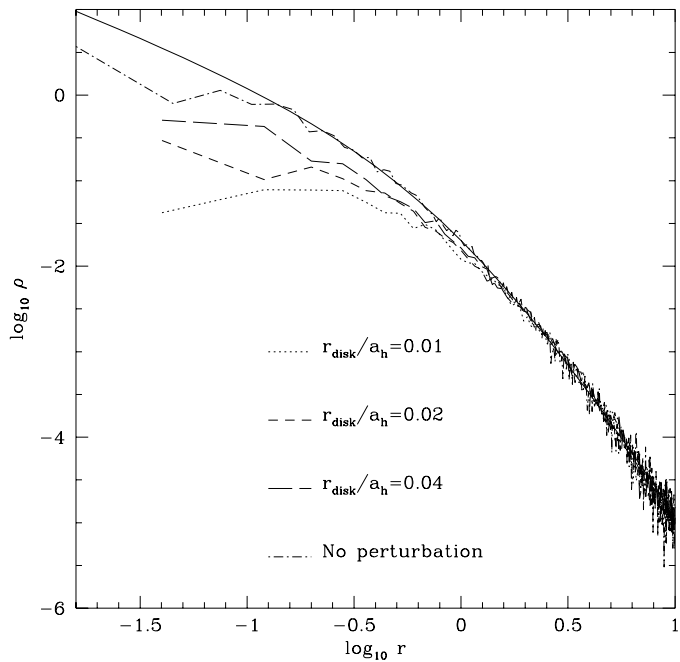

**Figure 3.** (b) *−continued.* $M_{disk} = 0.1$

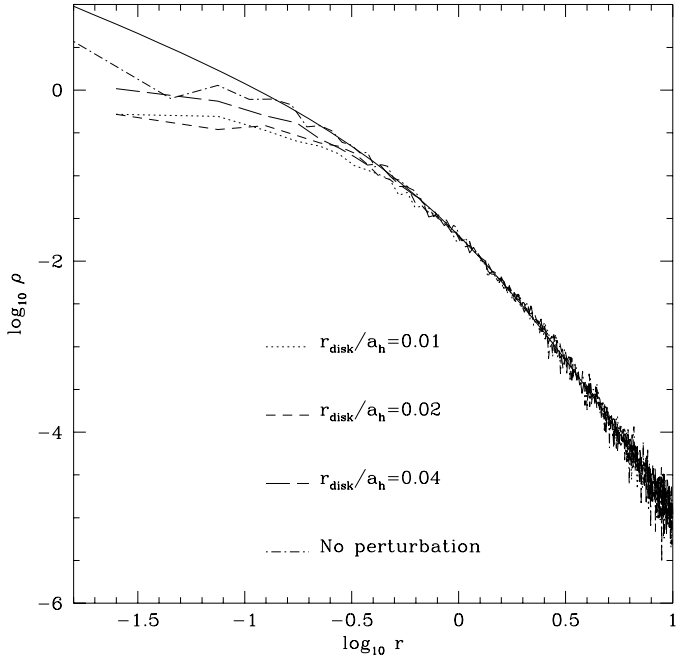

**Figure 3.** (c) –*continued*. $M_{disk} = 0.05$

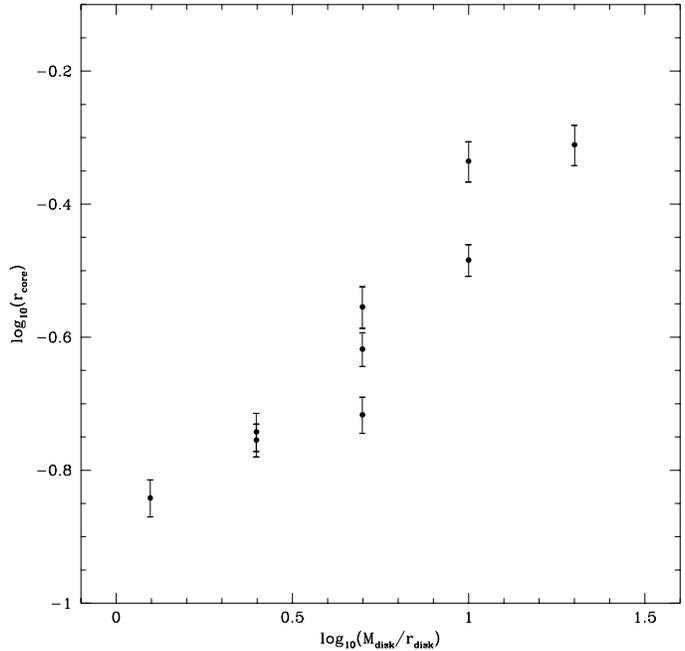

**Figure 4.** Core radii fitted to the profiles shown in Figure 3, as a function of the "strength" of the perturbation, $M_{disk}/r_{disk}$. The values of $r_{core}$ are in the computational units given in Section 2.1

the true contribution of the halo. The curves are shown in the radial range actually probed by the observations.

Matching these data to the rotation curves of the "perturbed" haloes shown in Figure 3 entails simply assigning physical units to the scale-length $a_h$ (in kpc) and $V_{max}$ (in km/s). We choose the values of these two parameters so that the rotation curves of the haloes fall between the two extreme cases allowed by the observations. The result is shown in Figure 5, where the rotation curves corresponding to the perturbed haloes are shown as symbols connected with dashed lines. In each panel, filled circles correspond to $r_{disk} = 0.04$, open circles to $r_{disk} = 0.02$, and starred symbols to $r_{disk} = 0.01$.

Figure 5 shows that all perturbed haloes can reproduce the observed structure of the haloes of dwarf irregulars, for suitably chosen values of $a_h$ and $V_{max}$. Note also that in some cases the maximum velocity of the halo is larger than the circular velocity at the outermost measured point. Figure 6 shows the values of $a_h$ and $V_{max}$ required to fit each galaxy for our nine models. Points in the upper left of this figure correspond to haloes that have suffered "strong" perturbations (large $M_{disk}/r_{disk}$), whereas those at the bottom right correspond to haloes that have been only "weakly" perturbed (small $M_{disk}/r_{disk}$). The points cover a wide range in $V_{max}$ and $a_h$, indicating that, depending on the strength of the perturbation, haloes of very different mass and size could be the progenitors of the haloes that we see today around dwarf galaxies.

## 4 DISCUSSION

Having established that mass outflows can modify the structure of a dark halo and imprint a core radius such as those seen in dwarf galaxies, we need to consider whether the disk masses and sizes required to produce a sizeable core are consistent with theoretical and observational constraints.

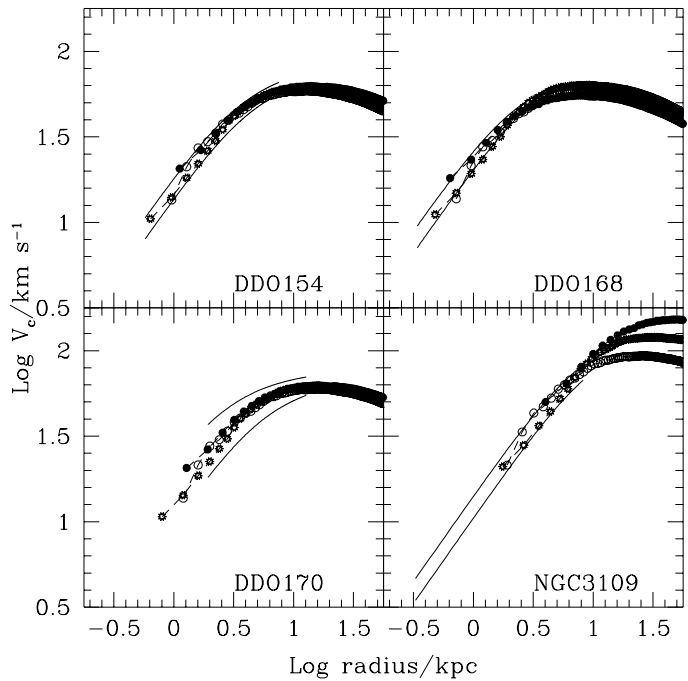

**Figure 5.** (a) Fits to the rotation curve of four dwarf galaxies ($M_{disk} = 0.2$). Different symbols correspond to different values of $r_{disk}$. The fits were made by assigning physical units to $a_h$ and $V_{max}$ for each perturbed halo so that their circular velocities lie between the two extreme cases allowed by observations (solid lines).

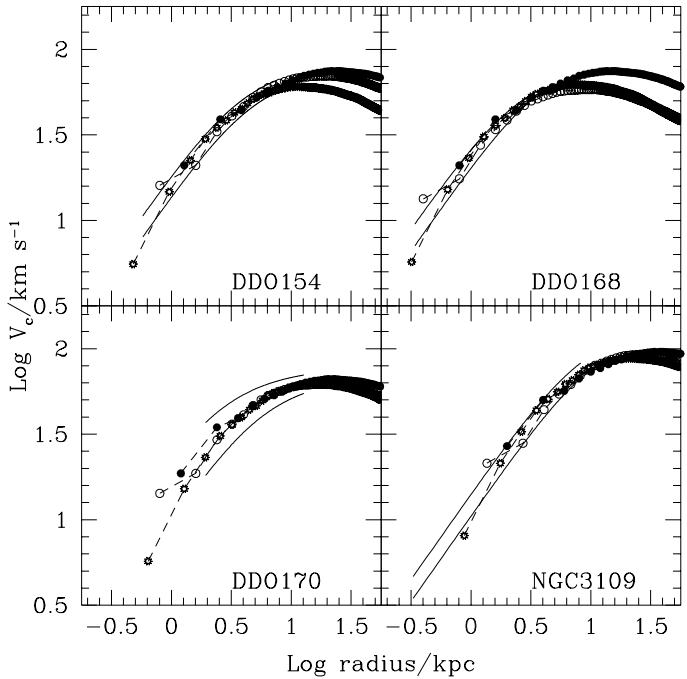

**Figure 5.** (b)–*continued.* $M_{disk} = 0.1$

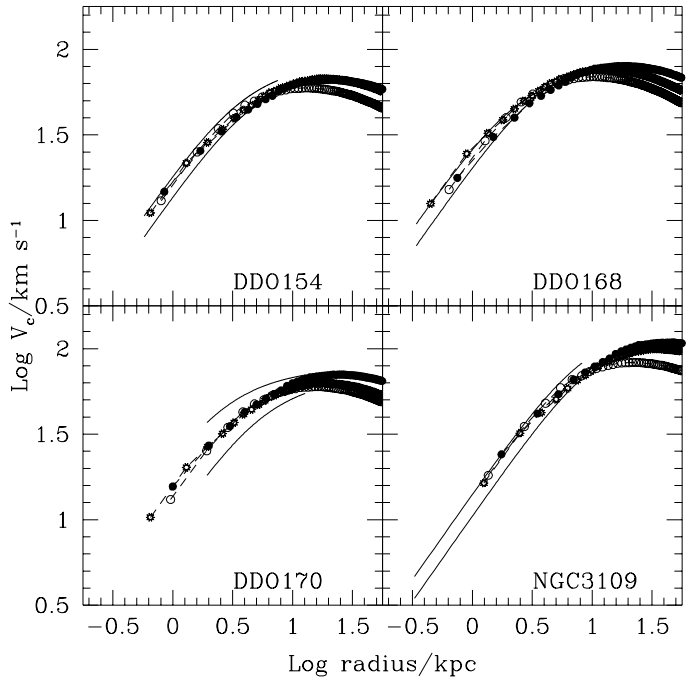

**Figure 5.** (c)–*continued.* $M_{disk} = 0.05$

An upper bound to the mass of a protogalactic disk follows from Big Bang nucleosynthesis constraints, if we assume that the primordial sites of dwarf galaxy formation had the universal mixture of baryonic and non-baryonic material and that all available baryonic material collapsed to the disk. The abundances of the light elements require the primordial baryon abundance to be $\Omega_b h^2 \lesssim 0.015$ (Walker *et al.* 1991). (We express the value of the Hubble constant as $100\,h$ km s$^{-1}$ Mpc$^{-1}$.) Since $h$ is unlikely to be less than 0.5, this implies a maximum baryon fraction of $\sim 6\%$ of the critical density. Assuming that $\Omega = 1$, as appropriate for the model we are testing, the standard CDM dominated universe, we conclude that the mass of the disk cannot be larger than 6% of the total mass (ie. $M_{disk} \lesssim 0.06$ in our computational units). This fraction would be smaller if some of the gas turned into stars before reaching the centre. To constrain the maximum possible perturbation on the halo we take the limiting case in which the mass fraction $\Omega_b$ condenses at the centre.

The typical size of the baryon disk can be constrained by angular momentum considerations. In a hierarchical clustering model like CDM the angular momentum of a galaxy originates from tidal torques exerted by neighbouring systems before turnaround. This is a rather inefficient process that imparts a halo (and its baryonic component) a net spin that may be quantified in terms of the dimensionless rotation parameter $\lambda = J|E|^{1/2}/GM^{5/2}$. (Here $J$ is the total angular momentum and $E$ is the binding energy.) Both numerical studies and theoretical work indicate that $\lambda \approx 0.05$, albeit with a fairly large scatter (Barnes and Efstathiou 1987).

Assuming that baryons conserve their angular momentum during the dissipative collapse that forms the disk, the exponential scale-length of the disk is expected to be not less than $\sim \lambda \times a_h$ (Fall & Efstathiou 1980) or, in our computational units, $r_{disk} \gtrsim 0.05$.

These simple considerations suggest that the ratio $M_{disk}/r_{disk}$ is unlikely to be much larger than unity, corresponding roughly to the weakest perturbations in our simulations. The core radius produced by such perturbations would be less than about 20% of the original scale-length of the halo (see Figure 4). Fitting the large cores observed in dwarf galaxies therefore requires dark matter haloes of a given circular velocity to have relatively large scale-lengths. According to Figure 6, for our models to be consistent with observations unperturbed haloes with $V_{max} \sim$ 50-100 km s$^{-1}$ should have scale-lengths as large as 20-30 kpc (see bottom right points on each panel in Figure 6).

Such large scale-lengths are consistent with the results of recent N-body simulations of the formation of Cold Dark Matter haloes (Navarro et al. 1996). These authors confirm that the inner regions of CDM haloes are very well approximated by a Hernquist model and present evidence that the scale-length and mass (or circular velocity) of a halo are strongly correlated. The correlation between $a_h$ and $V_{max}$ they find (see their Figure 10) is shown in Figure 6 with dotted lines. Except perhaps for DDO170, it is clear that the structure of slightly perturbed CDM haloes is consistent with the rotation curves of dwarf irregulars.

In our outflow model most of the baryonic disk is expelled on a timescale relatively short compared with the dynamical timescale of the system. Otherwise, the dark halo would respond adiabatically to the perturbation and no core would form. This imposes constraints on the magnitude and duration of the star formation burst that causes the outflow. The energy required to unbind a baryon mass $M_{gas}$ from a halo with circular velocity $V_c$ is of order $\sim M_{gas} V_c^2$. The total amount of energy dumped by supernovae is proportional to the total amount of stars formed during the burst. For a typical Miller-Scalo IMF, about 1 type II SN is expected to form per $100 M_\odot$ of stars formed, each dumping $10^{51}$ ergs into the interstellar medium. Assuming that a fraction $\epsilon_\star$ of the energy released by SN is invested into driving the outflow, and that the mass of gas to be blown out is just $\Omega_b$ times the total mass of the halo, the total mass of stars that need to form in order to expel the remaining baryonic component from a halo of circular velocity $V_c$ is $\approx (\Omega_b/\epsilon_\star)(V_c/\mathrm{km\ s^{-1}})^5 M_\odot$. (Here we have adopted $h = 0.5$ and assumed that the mass of a halo of circular velocity $V_c$ is $\sim 4.65 \times 10^5 (V_c/\mathrm{km\ s^{-1}})^3 M_\odot$, ie. the mass of an isothermal sphere with the given circular velocity, truncated at the radius where the mean interior overdensity is 200 times the mean density of the universe.)

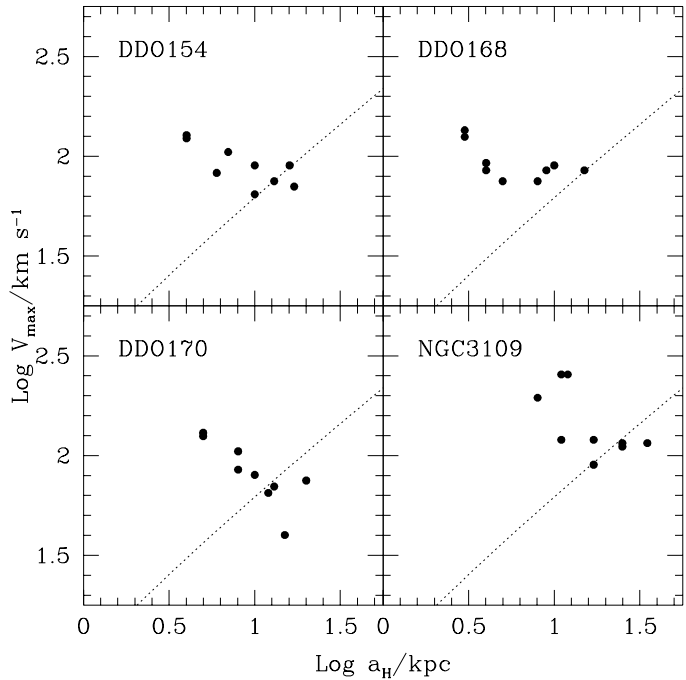

**Figure 6.** The values of the halo scale-length $a_h$ and maximum circular velocity $V_{max}$ chosen for the fits shown in Figure 5. The dotted lines correspond to the correlation reported by Navarro et al (1996) foor standard CDM halos.

This argument implies that $\sim 3 \times 10^8 (\Omega_b/\epsilon_\star) M_\odot$ need to form in order to blow out the baryonic component from a system with $V_c = 50$ km s$^{-1}$, a figure that compares well with the typical luminosity of these dwarf systems, roughly a few times $10^8 L_\odot$. To achieve this in a timescale shorter than the dynamical timescale within the luminous radius of the system, given roughly by $\sim 10$ kpc/50km s$^{-1} \approx 2 \times 10^8$ yrs, a star formation rate of about $\sim 1.5 (\Omega_b/\epsilon_\star) M_\odot$/yr is required. Assuming that $\Omega_b = 0.06$ (as required for consistency with primordial nucleosynthesis) and that only 1 % of the SN energy drives the wind (ie. $\epsilon_\star = 0.01$), a star formation rate of less than $10 M_\odot$/yr can produce the swift mass outflow envisaged in this simple model. Thus, the star formation rates required do not seem implausible. We conclude that massive rapid baryonic outflows may very well have played an important role during the formation of these systems. The steep dependence on circular velocity of the energy needed to trigger the outflow ($\propto V_c^5$) makes it unlikely that this mechanism would be efficient in more massive galaxies. Evidence for constant density cores in such galaxies may therefore need a different explanation. Studies of rotation curves of low surface brightness galaxies are particularly interesting, for in these systems the contribution of the luminous component to the overall rotation curve is smaller and thus the density structure of the dark halo is better constrained.

## 5 CONCLUSIONS

We have shown that sudden mass outflows can alter substantially the central structure of cold dark matter haloes. If a substantial fraction of the baryonic mass in a galaxy is blown out of the system by supernovae soon after it is assembled into a disk and before it is transformed into stars, it is possible to create a core in an originally coreless dark halo. The core radius thus created is a sensitive function of the mass and size of the disk. Disk masses and sizes consistent with nucleosynthesis constraints and angular momentum considerations can only imprint small cores in their surrounding haloes. This results in a strong upper bound for the central concentration of haloes of a given mass.

This upper bound is broadly consistent with the results of Navarro et al (1996) and imply that a rather weak perturbation can actually reconcile the structure of CDM haloes with the rotation curves of dwarf irregulars. We conclude that the shapes of the rotation curves of dwarf irregulars do not exclude the possibility that their haloes are made of cold dark matter. Rather, our argument emphasizes the role of feedback from an earlier generation of stars in removing baryonic material from the centre of the system and in devolving it to the surrounding halo. This process may explain the low baryon fraction in dwarf systems and has long been advocated for galaxies with even deeper potential wells than the dwarf irregulars we have considered here (Cole 1991, White & Frenk 1991, Cole et al 1994, Kauffmann et al 1994). Detailed rotation curve studies of a larger sample of galaxies, especially of low-surface brightness galaxies where, as in dwarf irregulars, the luminous component contributes a relatively minor fraction of the potential, would be extremely useful for setting constraints on the structure and composition of galactic dark matter haloes.